\documentclass[11pt]{article}
\usepackage[preprint]{acl}
\usepackage{times}
\usepackage{latexsym}
\usepackage[T1]{fontenc}
\usepackage[utf8]{inputenc}
\usepackage{microtype}
\usepackage{inconsolata}
\usepackage{graphicx}
\usepackage{subcaption}
\usepackage{xspace}
\usepackage{multirow}
\usepackage{booktabs}
\usepackage{colortbl}
\usepackage[dvipsnames]{xcolor}
\usepackage{tabularx}
\usepackage{tcolorbox}
\tcbuselibrary{skins,breakable}
\usepackage{xcolor}
\usepackage{pifont}
\usepackage{tcolorbox}
\tcbuselibrary{skins,breakable}
\usepackage[table]{xcolor}
\usepackage{pifont}
\usepackage{pgfplots}
\pgfplotsset{compat=1.18}
\usepackage{amsmath}
\usepackage{todonotes}

\newcommand{\modelname}{\textsc{RankVideo}\xspace}
\newcommand{\reasonrank}{\textsc{ReasonRank}\xspace}
\newcommand{\multivent}{\textsc{MultiVENT 2.0}\xspace}

\newcommand{\green}[1]{{\color{ForestGreen}#1}}
\newcommand{\red}[1]{{\color{BrickRed}#1}}

\newcommand{\cmark}{\textcolor{green!50!black}{\ding{51}}}
\newcommand{\xmark}{\textcolor{red!70!black}{\ding{55}}}

\title{\modelname: Reasoning Reranking for Text-to-Video Retrieval}

\author{
\textbf{Tyler Skow}\textsuperscript{\rm 1}\thanks{Equal Contribution}
\quad
\textbf{Alexander Martin}\textsuperscript{\rm 1}\footnotemark[1] \\
\quad \textbf{Benjamin Van Durme}\textsuperscript{\rm 1,2} 
\quad \textbf{Rama Chellappa}\textsuperscript{\rm 1}
\quad \textbf{Reno Kriz}\textsuperscript{\rm 1,2} 
\\
  \textsuperscript{1}Johns Hopkins University\quad \textsuperscript{2}Human Language Technology Center of Excellence\quad \\
  \texttt{\small{\{tskow1, amart233, rkriz1\}@jhu.edu}}
}

\begin{document}
\maketitle
\begin{abstract}

Reranking is a critical component of modern retrieval systems, which typically pair an efficient first-stage retriever with a more expressive model to refine results. While large reasoning models have driven rapid progress in text-centric reranking, reasoning-based reranking for video retrieval remains underexplored. To address this gap, we introduce \modelname, a reasoning-based reranker for video retrieval that explicitly reasons over query-video pairs using video content to assess relevance. \modelname is trained using a two-stage curriculum consisting of perception-grounded supervised fine-tuning followed by reranking training that combines pointwise, pairwise, and teacher confidence distillation objectives, and is supported by a data synthesis pipeline for constructing reasoning-intensive query-video pairs. Experiments on the large-scale \textsc{MultiVENT 2.0} benchmark demonstrate that \modelname consistently improves retrieval performance within a two-stage framework, yielding an average improvement of 31\% on nDCG@10 and outperforming text-only and vision-language reranking alternatives, while more efficient.\footnote{https://github.com/tskow99/RANKVIDEO-Reasoning-Reranker}
\end{abstract}

\section{Introduction}

Platforms across education, entertainment, and social media now host billions of videos, creating a growing demand for effective and scalable retrieval methods. Text-to-video retrieval (video retrieval) addresses this need by ranking large video collections in response to natural language queries. However, the task remains challenging due to the need for strong multimodal representations~\cite{samuel2025mmmorrfmultimodalmultilingualmodularized}, cross-modal alignment between text and audiovisual content~\cite{Reddy_2025_CVPR, ma2025tevatron20unifieddocument}, and the ability to scale to large real-world collections containing hundreds of thousands of videos~\cite{Kriz_2025_CVPR}.

A common paradigm in information retrieval (IR) for scalable systems is to pair an efficient bi-encoder~\cite{khattab2020colbertefficienteffectivepassage, warner2024smarterbetterfasterlonger} with a more expressive reranker that refines first-stage results~\cite{reimers2019sentencebertsentenceembeddingsusing, pradeep2023rankzephyreffectiverobustzeroshot}. This two-stage pipeline reduces the number of query-document comparisons, making the slower reranking tractable at query time. While recent work in video retrieval has produced strong first-stage models,~\cite{Reddy_2025_CVPR, samuel2025mmmorrfmultimodalmultilingualmodularized, faysse2025colpaliefficientdocumentretrieval, ma2025tevatron20unifieddocument, xu2025omniembednemotronunifiedmultimodalretrieval},\begin{figure}
    \centering
    \includegraphics[width=\linewidth]{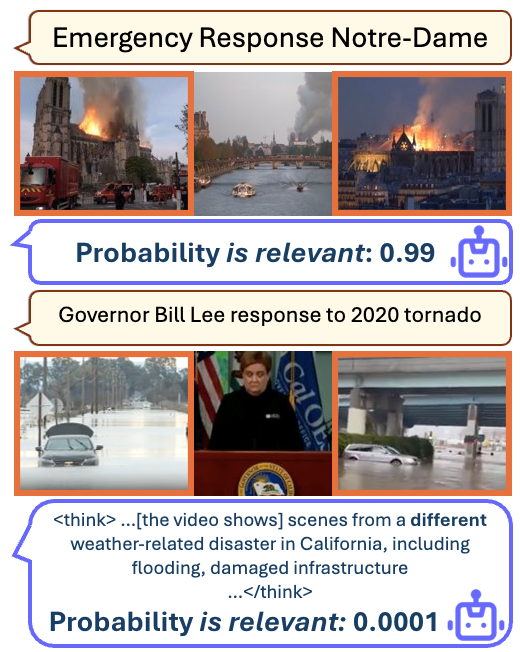}
    \caption{\modelname judges the relevance between a query-video pair, dynamically reasoning or answering depending on the difficulty of the query-video pair.}
    \label{fig:teaser}
    \vspace{-1em}
\end{figure} reranking remains largely unexplored. Leveraging textual reasoning models from IR is limited by the usefulness of extracted text (e.g., captions, transcribed speech, embedded text); this often omits critical visual or audio information, is not always available, and can be computationally expensive to generate.

In contrast, video-native reranking, which uses audiovisual inputs directly rather than relying on extracted text, provides a more robust alternative to text-only rerankers. Inspired by the recent success of Large Reasoning Models (LRMs) in multimodal understanding~\cite{li2025selfrewardingvisionlanguagemodelreasoning, bai2025qwen3vltechnicalreport, feng2025videor1reinforcingvideoreasoning} and reasoning reranking~\cite{weller2025rank1testtimecomputereranking, liu2025reasonrank, yang2025rankk}, we introduce \modelname, a video-native reasoning reranker for text-to-video retrieval. Given a query–video pair, \modelname predicts relevance by comparing the log-probabilities of discrete answer tokens, producing a scalar relevance score without requiring reasoning traces (\autoref{fig:teaser}).

To train an effective reranker, we first propose a data synthesis method for creating reasoning-intensive queries by leveraging the visual, audio, on-screen embedded text, and metadata features of videos. Using this data, we adopt a two-stage training process. In the first stage, perception-grounded supervised fine-tuning (SFT), the model learns to generate captions grounded in video content. In the second stage, the perception-grounded model is then fine-tuned as a reranker using a unified objective that combines pointwise classification, pairwise ranking, and teacher distillation. The pointwise term trains the model to classify query-video pairs as relevant or not, while the pairwise term encourages correct ranking of one positive video against two hard negatives per query. Soft-relevance probabilities are distilled from a large reasoning teacher, providing calibrated supervision that captures confidence beyond binary labels.

With this training process, \modelname achieves substantial retrieval gains across diverse first-stage retrievers, averaging 31\% improvement on nDCG@10, while remaining significantly faster than existing reasoning-based reranking baselines. We also observe that the model adaptively allocates reasoning effort, engaging in deeper reasoning only when necessary, which further improves efficiency without sacrificing performance.

Our contributions can be summarized as follows:
\begin{enumerate}
    \item We introduce \modelname, a video-native reasoning reranker for text-to-video retrieval trained with a two-stage curriculum directly on audiovisual inputs.
    \item We develop a data synthesis pipeline to generate reasoning-intensive query-video pairs. 
    \item We conduct extensive experiments demonstrating the effectiveness, generalizability, and efficiency of \modelname compared to strong text-only and vision-language reranking baselines.
\end{enumerate}

\section{Related}

\paragraph{Large Reasoning Models}
Large reasoning models (LRMs) extend pretrained language and multimodal models with the ability to perform multi-step reasoning, often by generating intermediate rationales or explicitly allocating additional computation at inference time \cite{bai2025qwen3vltechnicalreport, xu2025qwen3omnitechnicalreport, chen2025perceptionreasoningtwostagereinforcement, deepseekai2026deepseekr1incentivizingreasoningcapability}. Prior work has shown that structured reasoning with chain-of-thought or reinforcement learning can substantially improve performance on complex understanding and decision-making tasks \cite{wang2023selfconsistencyimproveschainthought, feng2025videor1reinforcingvideoreasoning, zhang2025consistentpathsleadtruth}. However, the amount of compute spent at test time is a common concern in the literature \cite{aggarwal2025l1controllinglongreasoning, cheng2024compressedchainthoughtefficient, hao2025traininglargelanguagemodels, yang2025thinkneedselfadaptivechainofthought}.

\paragraph{Reranking}
Neural information retrieval (IR) commonly adopts a two-stage approach to retrieval, seperating a fast, high-recall first-stage retriever from a more expressive reranker that operates on a samll subset of the first-stage results \cite{hui2017pacrrpositionawareneuralir, MacAvaney_2019_CEDR, MacAvaney_2019, pang2020setranklearningpermutationinvariantranking}. Canonically, reranking is performed using a cross-encoder that jointly encodes the query and each candidate document to produce a relevance score \cite{nogueira2019multistagedocumentrankingbert, reimers2019sentencebertsentenceembeddingsusing,nogueira2020passagererankingbert,  pradeep2023rankzephyreffectiverobustzeroshot}. Recently, large reasoning models have been adapted as rerankers, demonstrating substantial performance gains by explicitly scaling test-time compute for improved relevance judgments \cite{liu2025reasonrank,sun2025grouprankgroupwisererankingparadigm,weller2025rank1testtimecomputereranking,yang2025rankk, zhang-etal-2025-rearank}. While effective, existing approaches are largely developed for text-based retrieval. Our method follows this trajectory but extends reasoning-based reranking to a video-native setting, where relevance judgments require jointly reasoning over visual, audio, textual, and temporal signals.

\paragraph{Text-to-Video Retrieval}
Video retrieval is a core research area in video-language understanding \cite{yu2018jointsequencefusionmodel, yang2021tacotokenawarecascadecontrastive, wang-shi-2023-video, cao-etal-2024-rap, tang2025musemambaefficientmultiscale}. However, it has traditionally been done with captioning datasets converted to retrieval tasks at small scales \cite{chen-dolan-2011-collecting, hendricks2017msvd, krishna2017densecaptions, xu2017msrvtt, wang2019vatex}. \multivent \cite{Kriz_2025_CVPR} introduced a large-scale, more reasoning intensive dataset for video retrieval, which better reflected real world retrieval needs. \citet{Kriz_2025_CVPR} found that state-of-the-art methods don't scale to 100k+ videos or fail with real-world queries. Instead, for optimal performance and mirroring text IR, video retrieval should be split into two stages, where a first-stage retriever produces a ranked list on the entire index \cite{degenaro-etal-2025-fortify, ma2025tevatron20unifieddocument, Reddy_2025_CVPR, samuel2025mmmorrfmultimodalmultilingualmodularized, zhan2025magmarsharedtaskdescription} and then a subset of that list is reranked by a more expensive cross-encoder. This work looks to introduce the second stage of this process. Concurrent work also introduces a reranker for video content \cite{li2026qwen3vlembeddingqwen3vlrerankerunifiedframework}. While this model performs state-of-the-art on contrived retrieval tasks (e.g., MSR-VTT), we find it significantly decreases first stage performance in real-world retrieval settings.

\section{Data Synthesis}

Video retrieval lacks training data for extensive finetuning and challenging retrieval needs that better match real world video retrieval \cite[beyond descriptive captions turned into queries,][]{Kriz_2025_CVPR}. To create high-level, reasoning intensive queries, we first generate and extract text representations of the video content. We caption the videos with \textsc{Qwen3-Omni-30B-A3B-Instruct}~\cite{xu2025qwen3omnitechnicalreport}, transcribe the audio with \textsc{Whisper-Large-v2}~\cite{pmlrradford23a}, and extract OCR with a state-of-the-art multilingual OCR system~\cite{etter2023hybrid}. Using these texts as a proxy for the video content, we then provide a text reasoning model \cite[\textsc{Qwen3-32B},][]{yang2025qwen3technicalreport} 5 variations of the data: caption only, audio only, OCR only, metadata only, and all information.

We filter these queries to a high quality reasoning intensive subset and ensure the queries are not relevant to other videos (e.g., broad queries). To do this filtering, we first discard queries whose true positive video did not exist within the first 1000 candidates returned from \textsc{OmniEmbed} \cite{ma2025tevatron20unifieddocument}. Next, we removed queries whose top first hard negative had a first stage score more then 2x the size of the true positive. Finally, we discarded query video pairs wrongly classified by \reasonrank-32B, supplying video captions as evidence for judgment. Our filtered dataset contains 35684 records, with 9267 unique positive query-video pairs and 26258 negative query-video pairs. 
On average each query has 3.85 candidates: 7995 queries have 3 negative samples, 1014 queries have 2 negatives samples, 245 queries have 1  negative sample and 13 queries have 0 negative samples. 

\section{\modelname Two-Stage Training}

\begin{figure*}
    \centering
    \includegraphics[width=\linewidth]{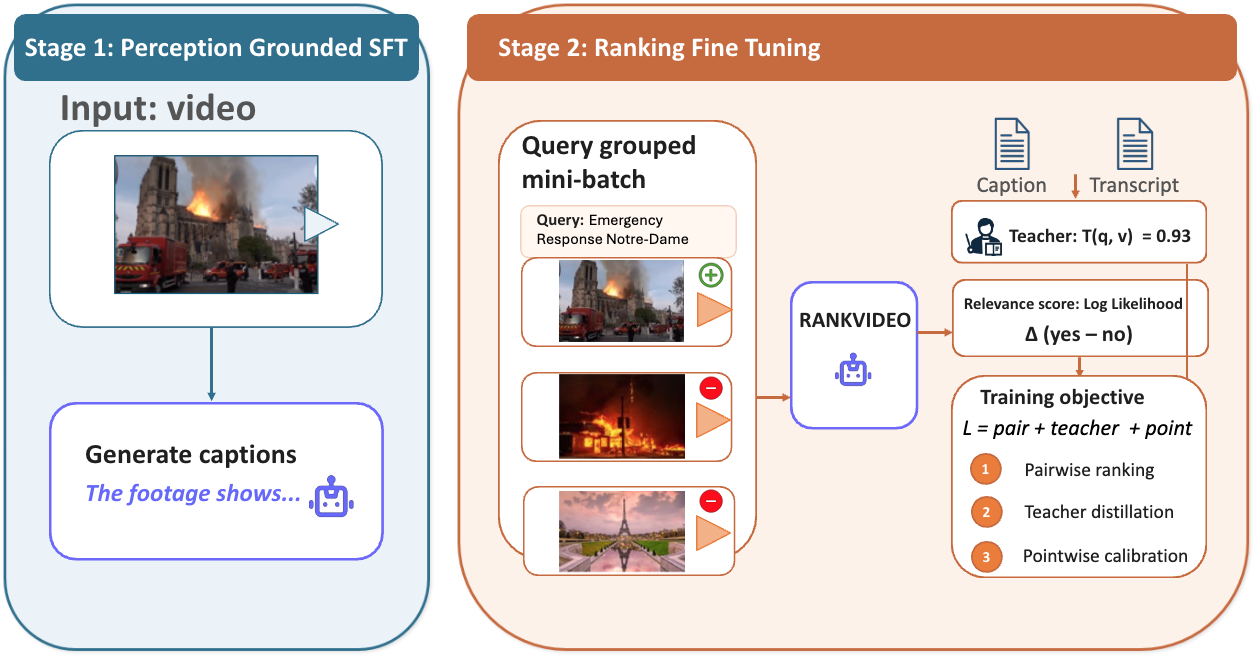}
    \caption{\modelname is trained with a two-stage process. Stage 1 uses a perception-grounded supervised finetuning, where the model learns to generate captions grounded in video content. In Stage 2, for each text query, we sample a query grouped batch containing one positive (relevant) video and one or more negatives (not relevant), and score each candidate using the difference between the logits for yes and no. The model is optimized with a combined objective: (1) teacher-probability distillation toward $p_{yes}$, (2) a pointwise loss for stable binary calibration, and (3) pairwise ranking loss that pushes the positive to the top within the query batch.}
    \label{fig:main}
    \vspace{-1em}
\end{figure*}
With our training data, a mixture of the human-written and synthetic queries, we are able to reasonably train a video-native reranker. Our training approach is composed of two stages: (1) perception grounded cold-start supervised SFT and (2) teacher guided ranking optimization.

\subsection{Stage 1: Perception Cold Start SFT}
An effective video native reranker must be able to reliably extract grounded evidence from raw video, including objects, actions and event context, and then align that evidence with a query. Motivated by recent findings that separating perception from downstream reasoning can improve multimodal training stability and performance \cite{chen2025perceptionreasoningtwostagereinforcement}, we introduce a perception grounded `cold start' stage before applying any ranking-specific supervision.

In this stage, we supervise the model to generate teacher provided captions for videos. Captioning provides a direct and dense learning signal for video understanding. Rather than only learning a binary relevance decision, the model is trained to produce an explicit textual description of salient entities present in the clip. Our intuition is that this encourages the model to attend to discriminative visual content. 

\paragraph{Training Data Construction}
We use teacher captions generated offline for the training videos. To bound compute and avoid repeatedly captioning near-duplicate candidates, we restrict this stage to one video per query so that the captioning objective covers a broad set of unique events rather than many candidates for the same query.

\paragraph{Objective}
Let $v$ denote a video and $c^{(T)} = (c^{(T)}_1,\dots,c^{(T)}_L)$ its teacher caption token sequence. We fine-tune the model parameters $\theta$ to maximize the likelihood of the teacher caption conditioned on the video. 
\begin{equation}
\mathcal{L}_{\mathrm{cap}}
\;=\;
-\sum_{t=1}^{L} \log p_{\theta}\!\left(c^{(T)}_{t}\mid c^{(T)}_{<t}, v\right)
\label{eq:caption_sft}
\end{equation}
 This stage produces a perception-grounded initialization that we then use as the starting point for Stage~2 ranking fine-tuning. Empirically, we find that this cold-start alone yields improvements.

\subsection{Stage 2: Ranking Finetuning}
Unlike stage 1, where the objective was detached from the video retrieval task, stage 2's aim is to directly improve the reranking ability of the model. There are two core components to the effectiveness of this training: hard negative mining and a three part training objective combining pointwise, pairwise, and teacher distillation.

\paragraph{Hard Negative Mining}
We want to find hard negatives to use in the pointwise portion of the training objective. We do this by partitioning our data into three categories: trusted negatives, suspected positives, and hard negatives, and keep trusted negatives and hard negatives to use during training.

For each query $q$, let $v^*(q)$ denote its labeled positive video within the first stage candidate pool. We treat the other candidate(s) $v\neq v^*(q)$ as a potential negative, but filter and stratify negatives using a reasoning teacher, \reasonrank. Concretely, the teacher provides a binary judgment, $\hat{y_T}(q,v) \in \{0,1\}$ and a confidence margin:
\[
\delta_t(q,v) = \ell_T(\text{yes} | q,v) - \ell_T(\text{no} | q,v)
\]
where $\ell_T$ is the teacher logit at the yes/no decision position. We partition non-positive candidates into: 
\begin{itemize}
    \item Trusted negatives, where $\hat{y_T}(q, v) =0$ and $\delta_T(q,v) \leq \alpha_1$
    \item Suspected positives, where $\hat{y_T}(q, v) =1$ with high margin. We use $\delta_T(q,v) >\alpha_2$. All samples that meet the $\alpha_2$ threshold are dropped to reduce false negative contamination. See \autoref{append:method} for details on threshold selection. 
    \item Hard negatives, consisting of the remaining candidates. We retain ambiguous negatives because they closely resemble the positive under the first stage retriever and thus dominate reranking errors, (see \autoref{append:disconnect}).
\end{itemize}

\paragraph{Training}
In Stage~2 we train the model to judge query video relevance using a three part objective: pointwise accuracy, pairwise ranking ability, and teacher distilled confidence. Given a query $q$ and a candidate video $v$, the model is prompted to answer \texttt{yes} or \texttt{no} using a structured output format \texttt{<answer>yes/no</answer>}. Rather than relying on generated text, we compute a scalar relevance score from the model's logits at the decision point: 
\begin{equation}
s_{\theta}(q,v)
\;=\;
\ell_{\theta}(\texttt{yes}\mid q,v)\;-\;\ell_{\theta}(\texttt{no}\mid q,v),
\label{eq:delta_score}
\end{equation}
where $\ell_{\theta}(t\mid q,v)$ denotes the model logit for token $t$ at the decision position after the \texttt{<answer>} tag. The logit delta score provides a stable, monotonic ranking signal, and enables fast scoring without decoding long rationales. 

Training proceeds on query grouped mini batches.
For each query $q$, we sample a candidate set
$\mathcal{B}_q = \{(q,v_i)\}_{i=1}^{K+1}$ containing one positive $v^+ = v^*(q)$ and $K$ negatives $\{v^-_1,\dots,v^-_K\}$
from the same query.\footnote{In our experiments we use $K=2$.} Now let $s_i = s_\theta(q,v_i)$ be the score for each candidate in $\mathcal{B}_q$.

\textbf{Pairwise ranking loss} We define a softmax distribution over candidates within the query batch: 
\begin{equation}
p_i \;=\; \frac{\exp(s_i/\tau_{pair})}{\sum_j \exp(s_j/\tau_{pair})},
\label{eq:softmax}
\end{equation}
and optimize a batch wise objective that pushes the positive to the top of its query group:
\[
L_{pair} = -\log p_+
\]
In \autoref{eq:softmax}, we use $\tau_{pair}$ to prevent early saturation and vanishing gradients as score gaps grow.\footnote{We use $\tau_{pair}=10$ in our experiments.} 
\textbf{Teacher probability distillation}
Next, we distill a teacher-provided relevance probability using a temperature-scaled binary cross-entropy with logits (BCEL) on the same score:
\begin{equation}
\mathcal{L}_{\mathrm{t}}
\;=\;
\mathrm{BCEL}\!\left(\frac{s_{\theta}(q,v)}{\tau_{\mathrm{teacher}}},\; p^{(T)}_{\mathrm{yes}}(q,v)\right),
\label{eq:teacher}
\end{equation}
This transfers calibrated confidence beyond the binary label and helps align scores across queries. 
\textbf{Pointwise loss} To stabilize training under class imbalance and provide pointwise supervision, we add a calibration loss with softened negative targets to account for noise in our training data.
Let $y\in\{0,1\}$ be the binary relevance label and define
\begin{equation}
\tilde{y} =
\begin{cases}
1.0 & \text{if } y = 1, \\
0.1 & \text{if } y = 0,
\end{cases}
\qquad\hspace{-2em}
w =
\begin{cases}
1.0 & \text{if } y = 1, \\
0.5 & \text{if } y = 0.
\end{cases}
\end{equation}
\begin{equation}
\mathcal{L}_{\mathrm{pt}}
\;=\;
\mathrm{BCEL}\!\left(\frac{s_{\theta}(q,v)}{\tau_{\mathrm{point}}},\; \tilde{y}; \text{weight} = w \right),
\label{eq:teacher}
\end{equation}
Combining \textbf{Pairwise ranking loss}, \textbf{Teacher probability distillation} and \textbf{Pointwise loss}, the final Stage~2 loss is a weighted sum \footnote{We set $\lambda_{teacher}=5$ and $\lambda_{pt} = 0.5$}:
\begin{equation}
\mathcal{L}
\;=\;
\mathcal{L}_{\mathrm{pair}}
\;+\;
\lambda_{\mathrm{teacher}}\,\mathcal{L}_{\mathrm{t}}
\;+\;\lambda_{pt}\mathcal{L}_{\mathrm{pt}}.
\label{eq:full_loss}
\end{equation}

\section{Experiments}
\paragraph{Evaluation Setup} 
We evaluate on the \multivent \cite{Kriz_2025_CVPR} test set, consisting of 109,800 videos. For evaluation, we retrieve the top 1000 candidates per query from the first stage retriever. We then score all candidates with the reranker report recall at 10 (R@10), 20 (R@20), 50 (R@50), and 100 (R@100), and normalized discounted cumulative gain (nDCG@N) for the same cutoffs as recall. For all metrics, a higher number indicates better performance.

\begin{table*}[]
\centering
\setlength{\tabcolsep}{4.7pt}
\begin{tabular}{l|cccccccc}
\toprule
\textbf{Method} &
\textbf{R@10} & \textbf{nDCG@10} &
\textbf{R@20} & \textbf{nDCG@20} &
\textbf{R@50} & \textbf{nDCG@50} &
\textbf{R@100} & \textbf{nDCG@100} \\
\midrule

\rowcolor{gray!30}
\multirow{1}{*}{OE}
 & 0.523 & 0.495 & 0.598 & 0.524 & 0.690 & 0.556 & 0.749 & 0.572 \\
 
\midrule

\multirow{2}{*}{RR}
 & 0.570 & 0.543 & 0.694 & 0.585 & 0.764 & 0.608 & 0.800 & 0.619 \\
 & \green{8.99} & \green{9.70} & \textbf{\green{16.05}} & \green{11.64} & \green{10.72} & \green{9.35} & \green{6.81} & \green{8.22} \\

\multirow{2}{*}{QVL-I}
 & 0.508 & 0.478 & 0.60 & 0.513 & 0.692 & 0.544 & 0.748 & 0.56 \\
 & \textbf{\red{-2.87}} & \red{-3.43} & \green{0.33} & \textbf{\red{-2.10}} & \green{0.29} & \textbf{\red{-2.16}} & \textbf{\red{-0.13}} & \red{-2.10} \\

\multirow{2}{*}{QVL-T}
 & 0.515 & 0.483 & 0.610 & 0.519 & 0.719 & 0.555 & 0.784 & 0.572 \\
 & \red{-1.53} & \red{-2.42} & \green{2.01} & \red{-0.95} & \green{4.20} & \red{-0.18} & \green{4.67} & N/A \\

\multirow{2}{*}{QVL-R}
 & 0.537 & 0.465 & 0.663 & 0.517 & 0.733 & 0.542 & 0.749 & 0.546 \\
 & \green{2.68} & \textbf{\red{-6.06}} & \green{10.87} & \red{-1.34} & \green{6.23} & \textbf{\red{-2.52}} & N/A & \textbf{\red{-4.55}} \\

\midrule 

\multirow{2}{*}{RV-1}
 & 0.582 & 0.559 & 0.661 & 0.589 & 0.727 & 0.612 & 0.752 & 0.619 \\
 & \green{11.28} & \green{12.93} & \green{10.54} & \green{12.40} & \green{5.36} & \green{10.07} & \green{0.40} & \green{8.21} \\

\multirow{2}{*}{RV-2}
 & 0.590 & 0.566 & 0.682 & 0.599 & 0.769 & 0.630 & 0.820 & 0.644 \\
 & \textbf{\green{12.81}} & \textbf{\green{14.34}} & \green{14.05} & \textbf{\green{14.31}} & \textbf{\green{11.45}} & \textbf{\green{13.31}} & \textbf{\green{9.48}} & \textbf{\green{12.59}} \\

\bottomrule
\end{tabular}
\caption{Performance changes from \textsc{OmniEmbed} first-stage retriever on \multivent. Each method reports raw scores (top) and deltas as a percentage relative to \textsc{OmniEmbed} (bottom). \green{Green} denotes an increase in performance, while \red{Red} denotes a decrease in performance. OE: \textsc{OmniEmbed}, RR: \textsc{ReasonRank}, QVL-I: \textsc{Qwen3-VL-8B-Instruct}, QVL-T: \textsc{Qwen3-VL-8B-Thinking}, QVL-R \textsc{Qwen3-VL-Reranker-8B}, RV-1/2: \modelname Stage 1/2.}
\label{tab:main_results}
\end{table*}

\paragraph{Results Setup}
Our main results are reranked from \textsc{OmniEmbed} \cite{ma2025tevatron20unifieddocument}, a dense retrieval model built on Qwen2.5 Omni \cite{xu2025qwen25omnitechnicalreport}, as the first-stage model. We also explore using four other first-stage systems: (1) \textsc{MMMORRF}~\cite{samuel2025mmmorrfmultimodalmultilingualmodularized} a multimodal rank fusion approach; (2) \textsc{CLIP}~\cite{radford2021learningtransferablevisualmodels} with 16 key frames selected by \textsc{PySceneDetect}~\cite{pyscenedetect}; (3) \textsc{LanguageBind}~\cite{zhu2024languagebindextendingvideolanguagepretraining} using connected language-vision encoders; (4) and \textsc{Video-ColBERT}~\cite{Reddy_2025_CVPR} a multi-vector late interaction model.  

For all first-stage retrievers, we rerank top 100 candidates using a first-stage depth of 1000. Along with \modelname, we evaluate four other baseline rerankers: (1)~\reasonrank \cite{liu2025reasonrank} a text-based reasoning reranker, which reranks the captions and audio transcripts of the video produced by \textsc{Qwen3-Omni-30B-A3B-Instruct}~\cite{bai2025qwen3vltechnicalreport} and \textsc{Whisper-Large-v2}~\cite{pmlrradford23a}; (2)~\textsc{Qwen3-VL-8B-Instruct} \cite[QVL-I,][]{bai2025qwen3vltechnicalreport}, a frontier video understanding model; (3)~\textsc{Qwen3-VL-8B-Thinking}~\cite[QVL-T,][]{bai2025qwen3vltechnicalreport}, the reasoning variant of QVL-I; and (4)~\textsc{Qwen3-VL-Reranker-8B}~\cite[QVL-R,][]{li2026qwen3vlembeddingqwen3vlrerankerunifiedframework}, the reranking variant of \textsc{Qwen3-VL}.

\paragraph{Implementation Details}
Our base model is intialized from \textsc{Qwen3-VL-8B-Instruct} \cite[QVL-I,][]{bai2025qwen3vltechnicalreport}. For stage one of training, we use one sample per query in order to generate a single unique caption per video, resulting in a training size set size of 9267. We fix the frame rate at 2 frames per second (FPS), with a maximum 32 frames per video, and train with a learning rate of 1e-5 and batch size of 16. In stage two of training, we form mini-batches of one positive and two negative queries. This yields a dataset of 7995 distinct queries with 23985 total videos. The final data mixture used to train our model contains 1361 human written queries and 7906 synthetic queries.

\subsection{\modelname and Baselines}
In \autoref{tab:main_results}, we show the results of the reranking methods on \multivent with \textsc{OmniEmbed} (OE) as the first-stage retriever. We see that \modelname achieves state-of-the-art retrieval results across all metrics, increasing the change in metric performance significantly more than any other reranking baseline. Both Stage-1 and Stage-2 of \modelname are able to significantly increase reranking performance, with the first and second best results across each metric. We find that the only other method to increase performance from the first stage results is \reasonrank, with strong gains over the first-stage results. The two other video-native baselines are not able to increase performance, struggling to judge the relevance of query-video pairs. QVL-I is unable to improve upon the first stage results. QVL-T is able to improve the recall from a cutoff of 20 and above, but unable to improve nDCG. This result means that QVL-T is able to remove easy negatives (lower in the ranked list), but struggles to rank relevant videos highly, especially against harder negatives in the first-stage results.

These results demonstrate three core findings. (1) Vision language or reasoning models (QVL-I/T) are not able to effectively judge the relevance of videos in zero-shot. This finding largely aligns with hallucination calibration literature \cite{li2025videohalluevaluatingmitigatingmultimodal, guan2024hallusionbenchadvanceddiagnosticsuite}, as we find QVL-I/T struggle with low precision as a result of a high false positive rate (QVL-I precision 0.055, QVL-T precision 0.037). (2) The gain in performance of \modelname over \reasonrank demonstrates the importance of performing \emph{video-native} reranking, incorporating heterogeneous multimodal signals into relevance judgments instead of extracting captions and transcripts, which are not information exhaustive. (3) Reranking models trained for contrived video-retrieval tasks\footnote{Video retrieval tasks that are captioning datasets converted to retrieval datasets} do not generalize well to real-world video retrieval \cite[mirroring the first stage findings of ][]{Kriz_2025_CVPR}. When comparing to QVL-R, we see the most significant decrease in performance from the first stage results, even compared to models not calibrated for reranking.
\begin{figure}
    \centering
    \includegraphics[width=\linewidth]{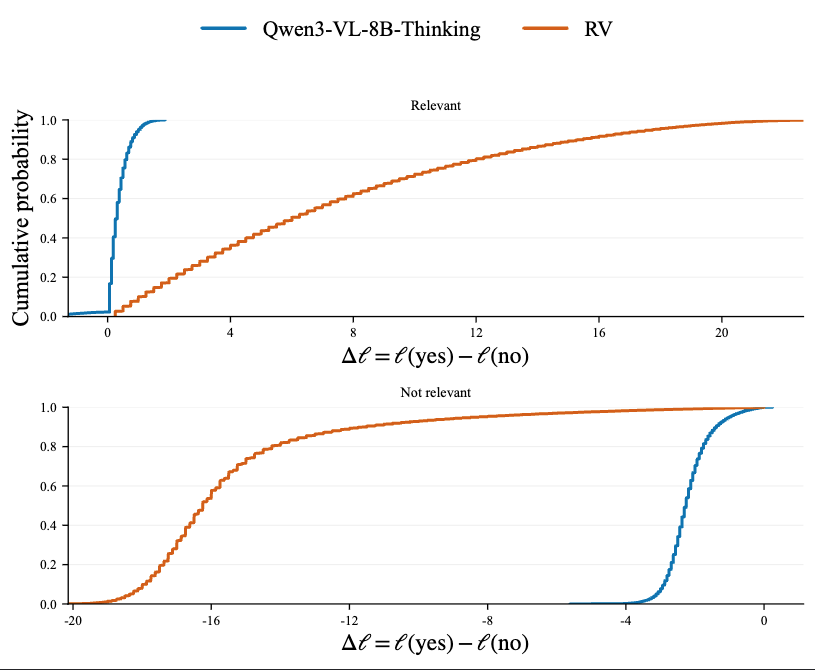}
    \caption{Training Stage-2 increases score separation in the reranking regime. Empirical CDF of the reranker score $s_{\theta}(q,v) = \ell_\theta(\textbf{yes}|q,v) - \ell_\theta(\textbf{no}|q,v)$ for relevant and non relevant query video pairs. Stage 2 shifts relevant pairs towards larger positive margins and suppresses non relevant candidates towards more negative margins, reducing overlaps in the score distributions within reranking candidate pools. }
    \label{fig:ecdf}
    \vspace{-1em}
\end{figure}
\subsection{Score Distribution Shift}
\modelname ranks candidates using logit delta score which provides a monotonic scalar ranking signal without decoding long rationales. Because reranking occurs over the first state top-1000 candidate pool, the dominant error mode is high scoring false positives among hard negatives. \autoref{fig:ecdf} shows that Stage 2 training reshapes the score distribution in a label consistent way: relevant pairs shift toward substantially larger positive margins, while non relevant candidates are pushed towards more negative margins, increasing separation and reduction overlap. This distribution shift is key for improving rank quality, especially in early ranks, since top-k metrics are most sensitive to the impact of high scoring hard negatives outranking the true item. This results suggest training does not just improve point wise correctness, but also calibrates the scores to match the reranking objective. Additionally, the increased separation suggests that the logit margin score can serve as a more reliable confidence signal. 

\begin{table*}[t]
\centering
\setlength{\tabcolsep}{4pt}
\begin{tabular}{l|cccccccc}
\toprule
\textbf{Method} &
\textbf{R@10} & \textbf{nDCG@10} &
\textbf{R@20} & \textbf{nDCG@20} &
\textbf{R@50} & \textbf{nDCG@50} &
\textbf{R@100} & \textbf{nDCG@100}
\\
\midrule

\rowcolor{gray!30}
\multirow{1}{*}{CLIP}
& 0.333 & 0.306 & 0.419 & 0.339 & 0.522 & 0.373 & 0.603 & 0.394 \\

\multirow{2}{*}{+ RV}
& 0.477 & 0.478 & 0.540 & 0.503 & 0.584 & 0.520 & 0.603 & 0.525 \\
& \green{43.24} & \green{56.21} & \green{28.88} & \green{48.38} & \green{11.88} & \green{39.41} & N/A & \green{33.25} \\

\midrule

\rowcolor{gray!30}
\multirow{1}{*}{MRF}
& 0.611 & 0.585 & 0.697 & 0.620 & 0.779 & 0.649 & 0.828 & 0.663 \\

\multirow{2}{*}{+ RV}
& 0.634 & 0.639 & 0.725 & 0.639 & 0.800 & 0.665 & 0.828 & 0.673 \\
& \green{3.76} & \green{8.45} & \green{4.02} & \green{3.06} & \green{2.70} & \green{2.47} & N/A & \green{1.51}\\

\midrule

\rowcolor{gray!30}
\multirow{1}{*}{LB}
& 0.355 & 0.326 & 0.445 & 0.359 & 0.550 & 0.393 & 0.620 & 0.412 \\

\multirow{2}{*}{+ RV}
& 0.498 & 0.487 & 0.560 & 0.512 & 0.607 & 0.530 & 0.620 & 0.533 \\
& \green{40.28} & \green{49.39} & \green{25.84} & \green{42.62} & \green{10.36} & \green{34.86} & N/A & \green{28.37} \\

\midrule

\rowcolor{gray!30}
\multirow{1}{*}{VCB}
& 0.341 & 0.422 & 0.402 & 0.447 & 0.471 & 0.477 & 0.518 & 0.494 \\

\multirow{2}{*}{+ RV}
& 0.448 & 0.535 & 0.490 & 0.549 & 0.514 & 0.560 & 0.518 & 0.561 \\
& \green{31.38} & \green{26.78} & \green{21.89} & \green{22.82} & \green{9.13} & \green{17.40} & N/A & \green{13.56} \\

\bottomrule
\end{tabular}
\caption{Impact of \modelname applied to different first-stage retrievers on \multivent. Each method reports raw retrieval performance (top) and absolute deltas as a percentage relative to its corresponding baseline retriever (bottom). \green{Green} indicates an improvement in ranked list quality. CLIP: CLIP on 16 frames, MRF: MMMORRF, LB: LanguageBind, VCB: Video-ColBERT, RV: \modelname.}
\label{tab:stability_results}
\vspace{-1em}
\end{table*}
\subsection{Stability Across First-Stage Model}
In \autoref{tab:stability_results} we explore the stability of \modelname across the other first-stage retrieval methods on \multivent. Like OmniEmbed, we observe a significant gain in performance from first-stage to second-stage results for all first-stage methods. On a strong first stage retriever (MMMORRF), we continue to see solid gains between ranked lists. Most promisingly, we see the largest and most significant gains (>0.1) on the weakest first-stage retrievers.

\begin{figure}
    \centering
    \includegraphics[width=\linewidth]{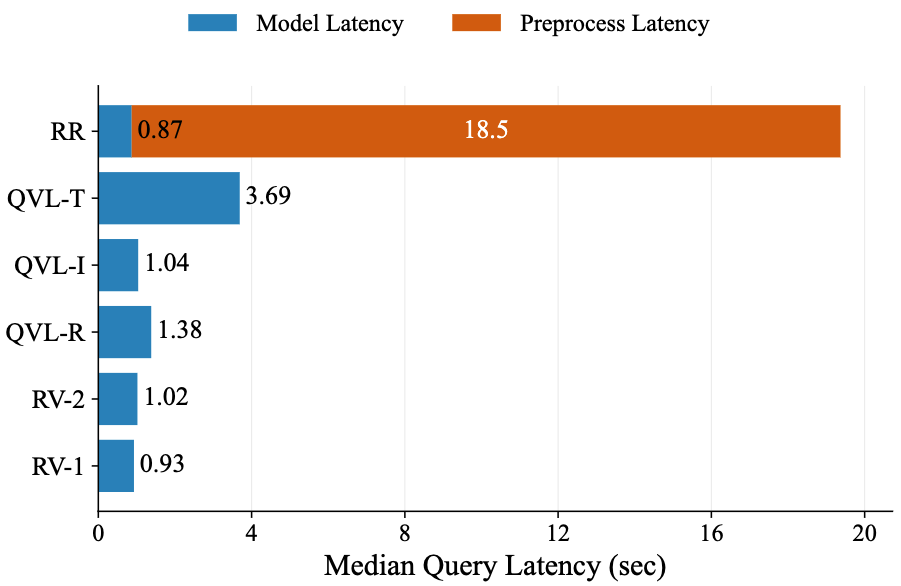}
    \vspace{-2em}
    \caption{Median query latency for Qwen3VL Instruct/Thinking (QVL-I/T) and \modelname stages 1 and 2. Latency is computed as the mean for 100 query-video pairs with a batch size of 1. All evaluations are run with batch size 1 as larger batches exceed GPU memory for VLMs.}
    \label{fig:latency_median_bar}
\end{figure}

These results demonstrate that \modelname is beneficial to any first-stage retriever. Additionally, \modelname is not dependent on the quality of the initial candidate list, allowing for the use of faster, more efficient (although less accurate) first-stage models to handle large indices and rely on \modelname to refine the ranked list. 

\subsection{Efficiency and Dynamic Reasoning}
In \autoref{fig:latency_median_bar} we compare the query latency of ReasonRank, QVL-T/I/R and \modelname Stages 1 and 2 (RV-1/2) on a randomly selected subset of 100 query-video pairs from \multivent, reporting the median query latency on these 100 instances.\footnote{We chose 100 instances to match the reranking cutoff.} We find that \modelname is much more efficient than the alternative baseline video-native reasoning model (QVL-T), with a difference of 2.67s between \modelname Stage 2 and QVL-T. We attribute the large difference in query latency because QVL-T produces a reasoning trace for every query-video pair. 

In \autoref{fig:latency_median_bar}, we observe that \modelname performs within 0.15s of ReasonRank (1.02s vs. 0.87s). We include a preprocessing latency penalty for ReasonRank to contextualize  the offline compute costs for captioning. While our model does require captions for training, at inference, \modelname's reliance soley on video frames avoids expensive preprocessing latency. This comes at the trade off of only a small gap in model latency vs. text based rerankers (ReasonRank). ReasonRank's latency values relative to all other models should be interpreted as an amortized per candidate estimate, since ReasonRank is listwise and evaluates a sliding window of candidates simultaneously.

\begin{table}[t]
\setlength{\tabcolsep}{4pt}
\centering
\begin{tabular}{lccc|c}
\toprule
\textbf{Method} & $\boldsymbol{\alpha}$-\textbf{nDCG} & \textbf{nDCG} & \textbf{StRecall} & \textbf{InfoP} \\
\midrule
    CLIP & 0.538 & 0.498 & 0.724 & 81.5 \\
    +RV   & \textbf{0.628} & \textbf{0.629} & \textbf{0.826} &\textbf{91.9} \\
\midrule
    LB   & 0.476 & 0.457 & 0.634 & 84.0 \\
    +RV   & \textbf{0.566} & \textbf{0.563} & \textbf{0.754} & \textbf{94.5} \\
\midrule
    VCB  & 0.431 & 0.383 & 0.634 & 83.8 \\
    +RV   & \textbf{0.553} & \textbf{0.522} & \textbf{0.734} & \textbf{87.1} \\
\midrule
    OE   & 0.530 & 0.454 & 0.721 & 88.0 \\
    +RV   & \textbf{0.584} & \textbf{0.587} & \textbf{0.778} & \textbf{91.2} \\
\midrule
    MRF  & 0.540 & 0.503 & 0.724 & \textbf{94.4} \\
    +RV   & \textbf{0.605} & \textbf{0.617} & \textbf{0.785} & 91.8 \\
\bottomrule
\end{tabular}
\caption{Retrieval and generation results on \textsc{WikiVideo}. The retrieval results are calculated using claim-based relevance and a cutoff at 10.}
\label{tab:rag_results}
\vspace{-1em}
\end{table}

\subsection{Downstream Impact in RAG}
We explore the effect of \modelname in a retrieval augmented generation setting. We perform RAG with the \textsc{WikiVideo} dataset \cite{martin2025wikivideoarticlegenerationmultiple}, evaluating retrieval with three claim-based relevance judgments $\alpha$-nDCG, nDCG, and StRecall, where the number of supported article claims are used for document relevance; evaluating retrieval with \textsc{MiRAGE} \cite{martin2025seeingmirageevaluatingmultimodal}, reporting Information Precision (InfoP) for article factuality; and generating articles using CAG with a \textsc{Qwen-3-VL} backbone \cite{bai2025qwen3vltechnicalreport, martin2025wikivideoarticlegenerationmultiple}.  In \autoref{tab:rag_results}, we find that \modelname substantially improves upon the claim coverage of the top 10 videos provided to the generation system, which leads to a large increase in article factuality. This demonstrates the effectiveness of \modelname, not only as an effective reranker for traditional retrieval metrics, but also as a crucial step in a multimodal RAG pipeline, increasing diversity of information for generation.

\definecolor{ModelBlue}{RGB}{30,130,181}         
\definecolor{PreprocessOrange}{RGB}{212,87,27}    
\newcolumntype{Y}{>{\raggedright\arraybackslash}X}
\begin{table}[t]
\centering
\small
\setlength{\tabcolsep}{6pt}
\renewcommand{\arraystretch}{1.10}

\begin{tabularx}{\linewidth}{@{}Yccc@{}}
\toprule
\textbf{Query} & \textbf{Before $\downarrow$} & \textbf{After $\downarrow$} & \textbf{$\Delta \downarrow$} \\
\midrule
KAMAZ Autonomous Mining Dump Truck                 & 100 & 1  & \cellcolor{ModelBlue!35}{-99} \\
Korea Blockchain Week 2022                         & 98  & 5  & \cellcolor{ModelBlue!33}{-93} \\
2018 Mark Zuckerberg senate hearing                & 92  & 1  & \cellcolor{ModelBlue!32}{-91} \\
Sixth SpaceX operational mission                   & 93  & 2  & \cellcolor{ModelBlue!32}{-91} \\
2019 Cairo International Film Festival             & 2   & 1  & -1 \\
2019 Moscow International Film Festival            & 1   & 1  & 0 \\
Digital Services Act                               & 1   & 1  & 0 \\
2019 Typhoon Lekima                                & 2   & 5  & \cellcolor{PreprocessOrange!12}{+3} \\
Typhoon Maysak                                     & 2   & 6  & \cellcolor{PreprocessOrange!15}{+4} \\
2018 opinion rigging scandal in South Korea        & 11  & 44 & \cellcolor{PreprocessOrange!30}{+33} \\
\bottomrule
\end{tabularx}

\caption{Most improved and degraded queries after reranking.
\textbf{Before:} rank without reranking.
\textbf{After:} rank with reranking.
$\Delta$: change in rank (After $-$ Before); lower is better. The maximum rank is capped at 100.}
\label{tab:qualitative}
\vspace{-1.5em}
\end{table}
\subsection{Qualitative Results Discussion}

We analyzed the failure modes of \modelname by examining query performance grouped by metadata attributes. Overall, we observe relatively stable performance across, event types, modalities and some performance stratification across languages (see \autoref{append:query-types}).

To quantify whether coarse metadata can explain which queries the reranker performs poorly on, we fit a regression model to predict per query \textit{nDCG@10} using \textit{video language}, \textit{query event type}, \textit{video type}, \textit{video modality}, and \textit{query length}. We randomly split queries into train (75\%) and test (25\%) and train a Random Forest Regressor. On held out queries, the model achieves $R^2 = 0.093$, indicating that these coarse metadata features explain only a small fraction of variance in per query \textit{nDCG@10}. Feature importances suggest query length is the most informative single feature, consistent with the intuition that shorter/underspecified queries are more difficult.

Separately, we test whether \modelname’s ranking scores exhibit query or video specific priors by decomposing variance in the reranker score $s(q,v)$. Let $N_q$ denote the number of scored candidates for query $q$, and define the query mean score $\mu_q = \frac{1}{N_q}\sum_{v} s(q,v)$. Let $N_v$ denote the number of queries for which video $v$ is scored, and define the per video mean score $\mu_v = \frac{1}{N_v}\sum_{q} s(q,v)$. We evaluate three simple predictors of $s(q,v)$: (i)~query only $\hat{s}(q,v)=\mu_q$, (ii) video only $\hat{s}(q,v)=\mu_v$, and (iii) additive $\hat{s}(q,v)=\mu_q+\mu_v-\mu$ where $\mu$ is the global mean. The resulting $R^2$ values are 0.139 (query only) and 0.090 (video only), indicating that \modelname~’s scoring is not dominated by a video level prior and instead depends substantially on query--video interaction. In contrast, baseline rerankers exhibit stronger video priors (e.g., \textsc{QVL-R}: $R^2=0.755$). In \autoref{tab:r2_priors} we see the broader trend that the weaker performing re-ranker also appear to have strong modality bias.

Qualitatively, we attribute the performance of \modelname to its effectiveness with videos and queries that have visually anchorable events. Consider, for example, the queries that see that greatest improvements in \autoref{tab:qualitative}. Mining dump trucks and SpaceX operational mission are object specific and have discrete visuals (trucks and rockets).  On the other side, we attribute the negative performance of the other vision baselines to non-visual or weakly visual topics, like "2018 opinion rigging scandal in South Korea."  One confounding category of events the model appears to occasionally struggle with, despite intuition suggesting their ought to be an abundance of distinct visually grounded signals, is natural disasters. A plausible explanation is, storm footage, for example, may share more generic visuals (rushing water) rather than unique cues needed for quality reranking, leading to negative results. 
\begin{table}[t]
\centering
\small
\setlength{\tabcolsep}{6pt}
\begin{tabular}{l|ccc}
\toprule
\textbf{Method} & \textbf{Query Only} & \textbf{Video Only} & \textbf{Q+V} \\
\midrule
QVL-I & 0.192 & 0.119 & 0.262 \\
RR    & 0.033 & 0.105 & 0.130 \\
QVL-T  & \underline{0.295} & \underline{0.265} & \underline{0.431} \\
QVL-R  & 0.108 & \underline{0.755} & \underline{0.739} \\
\midrule
RV & 0.139 & 0.090 & 0.200 \\
\bottomrule
\end{tabular}
\caption{Query/video prior strength via variance decomposition. Stronger biases underlined.}
\label{tab:r2_priors}
\vspace{-1.52em}
\end{table}

\section{Conclusion}

In this work, we introduce \modelname, a novel approach for video-native reasoning reranking. \modelname is trained with a two stage process. The first stage perception-grounded SFT teaches the model to generate captions grounded in video content, while the second stage finetunes the stage 1 model for effective reranking by combining pairwise, pointwise, and distillation objectives. On \multivent, a large scale video retrieval task, \modelname consistently enhances retrieval performance across various first-stage retrievers, achieving an average improvement of 31\% on nDCG@10. Not only is \modelname the most effective reranker, but it is also significantly faster than existing reasoning-based reranking baselines. Future work in video reranking should look towards new training objectives, like list-wise or grouped reranking which exhibit strong performance gains over their pointwise counterparts in text IR \cite{liu2025reasonrank, sun2025grouprankgroupwisererankingparadigm, yang2025rankk}. Additionally, exploring further optimizations for dynamic reasoning, the disconnect between accuracy and retrieval performance, and improving performance on reasoning-intensive or ambiguous video types suggest ample direction for future work. In applications, reranking and modified objectives could be explored for retrieval-augmented generation to provide better optimized results to article generation models.

\section*{Limitations}
\paragraph{Computational Costs}
In this work, we did not explore list-wise reranking, even though it's benefit over pointwise is well supported in the literature, because of the computational costs of multivideo inference. To train our pairwise objective (at most 3 videos per query), required greatly reducing the batch size and max frames to fit on 8 80GB A100s for training. Efforts towards making multi-video inference more computationally feasible will help reduce this burden.

\section*{Acknowledgments}
This material is based upon work supported by the National Science Foundation Graduate Research Fellowship under Grant No. DGE2139757. Any opinion, findings, and conclusions or recommendations expressed in this material are those of the author(s) and do not necessarily reflect the views of the National Science Foundation.
\bibliography{custom}

\appendix

\section{\modelname Details}
\label{append:method}

We provide additional details about our training configurations in \autoref{tab:hyperparams} and provide the system and user prompts for \modelname (\autoref{prompt:stage1_prompt}, \autoref{prompt:method_prompt}), QVL-I/T (\autoref{prompt:qvl}, same as \modelname), and ReasonRank (\autoref{prompt:reasonrank}). For hard negative mining, we used thresholds $\alpha_1=-6$ and $\alpha_2 = -8$ after reviewing the distribution of logit scores of all query, video pairs.

\begin{table}[h]
    \centering
    \begin{tabular}{c|c}
    \toprule
        Setting & Value \\
    \midrule
        Learning Rate Schedule & cosine \\
        Learning Rate & 1e-5\\
        Warmup Proportion (Linear) & 0.03 \\
        Optimizer & AdamW \\
    \midrule
        Frame rate (FPS) & 2 \\
    \midrule
        Stage 1 Data & 9267/9267 \\
        Stage 1 Batch Size & 16 \\
        Stage 1 Epochs & 1 \\ 
        Stage 1 Max Frames & 32 \\
        Stage 2 Data & 7995/23985 \\
        Stage 2 batch Size & 3 \\ 
        Stage 2 Epochs & 2 \\
        Stage 2 Max Frames & 24 \\
    \bottomrule
    \end{tabular}
    \caption{Training settings for \modelname. Stage 1/2 Data is written as queries/videos. }
    \label{tab:hyperparams}
\end{table}

\section{Training Loss Ablation}
\label{append:training-loss-ablation}
\begin{table*}[t]
\centering
\setlength{\tabcolsep}{4pt}
\begin{tabular}{l|cccccccc}
\toprule
\textbf{Method} &
\textbf{R@10} & \textbf{nDCG@10} &
\textbf{R@20} & \textbf{nDCG@20} &
\textbf{R@50} & \textbf{nDCG@50} &
\textbf{R@100} & \textbf{nDCG@100}
\\
\midrule

\rowcolor{gray!30}
\multirow{1}{*}{OE}
& 0.554 & 0.541 & 0.614 & 0.566 & 0.711 & 0.599 & 0.755 & 0.614 \\

\multirow{1}{*}{P, PT, T}
& 0.650 & 0.640 & 0.694 & 0.657 & 0.731 & 0.672 & 0.755 & 0.678 \\

\multirow{1}{*}{P, PT}
& 0.650 & 0.630 & 0.687 & 0.644 & 0.733 & 0.663 & 0.755 & 0.668 \\

\multirow{1}{*}{P}
& 0.630 & 0.620 & 0.668 & 0.634 & 0.705 & 0.649 & 0.755 & 0.661 \\

\bottomrule
\end{tabular}
\caption{Performance changes from \textsc{OmniEmbed} first-stage retriever on \multivent. OE: \textsc{OmniEmbed}; P, PT, T: Our full loss objective; P, PT: Our loss objective without teacher distillation; P: Pairwise loss.}
\label{tab:loss-ablation}
\end{table*}

We include the ablation results of our three part loss objective in \autoref{tab:loss-ablation} to understand which signals are responsible for reranking gains. This ablation includes our pairwsie ranking loss (P), pointiwse calibration loss (PT), and teacher probability distillation (T).  We report results on a 50 query subset of the \textsc{MultiVENT 2.0}  test set. We observe substantial gains from the inclusion of the pointwise objective. The inclusion of the teacher probability distillation term resulted in an improved nDCG.

\section{\modelname Performance Across Query Types}
\label{append:query-types}

\begin{figure}[t]
  \centering
  \includegraphics[width=0.99\linewidth]{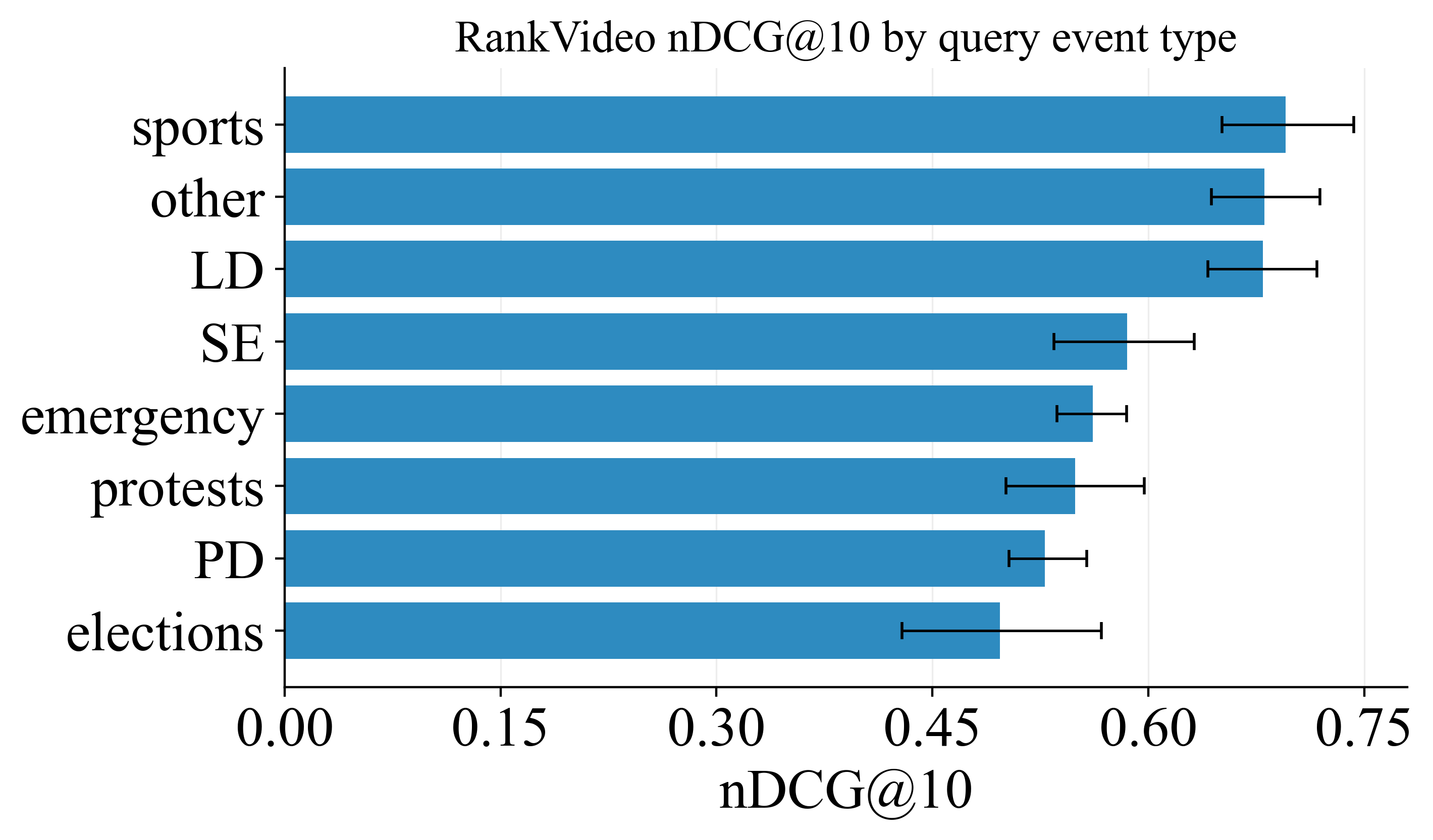}
  \caption{\modelname nDCG@10 by query event type. Multi-word categories are abbreviated as acronyms in the plot (e.g., PD=Political Development, LD=Launch/Discovery, SE=Social Events). Only attributes with $\ge$ 30 test queries are included.}
  \label{fig:rv-ndcg-event}
\end{figure}

\begin{figure}[t]
  \centering
  \includegraphics[width=0.99\linewidth]{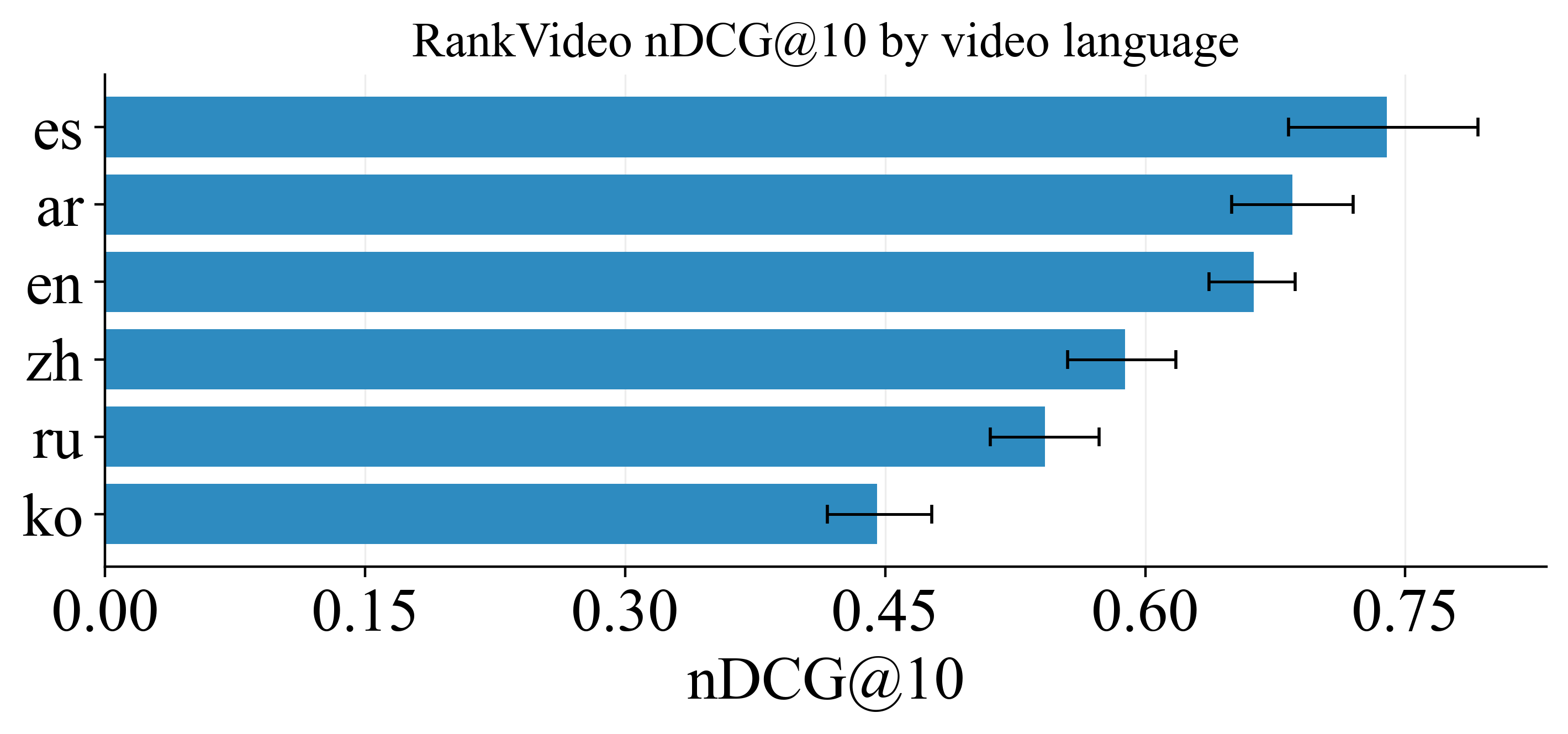}
  \caption{\modelname nDCG@10 by video language (en=English, es=Spanish, ar=Arabic, zh=Chinese, ru=Russian, ko=Korean). Only attributes with $\ge$ 30 test queries are included.}
  \label{fig:rv-ndcg-lang}
\end{figure}

\begin{figure}[t]
  \centering
  \includegraphics[width=0.99\linewidth]{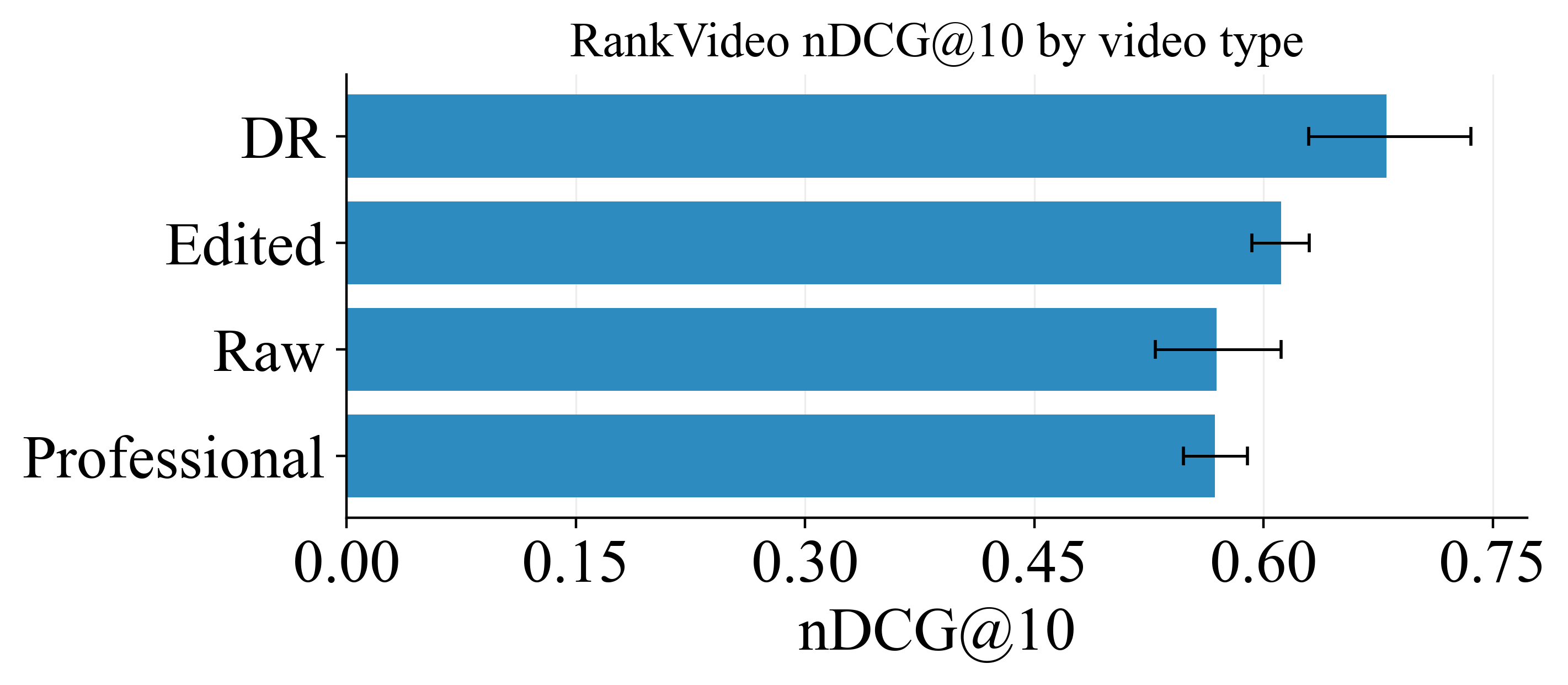}
  \caption{\modelname nDCG@10 by video type. Professional: e.g., news broadcasts with reports; Edited: e.g., videos with multiple spliced clips and visual effects; Diet Raw (DR): single-stream videos with minimal text/speech overlays; Raw: continuous, unedited streams. Only attributes with $\ge$ 30 test queries are included.}
  \label{fig:rv-ndcg-vtype}
\end{figure}

\begin{figure}[t]
  \centering
  \includegraphics[width=0.99\linewidth]{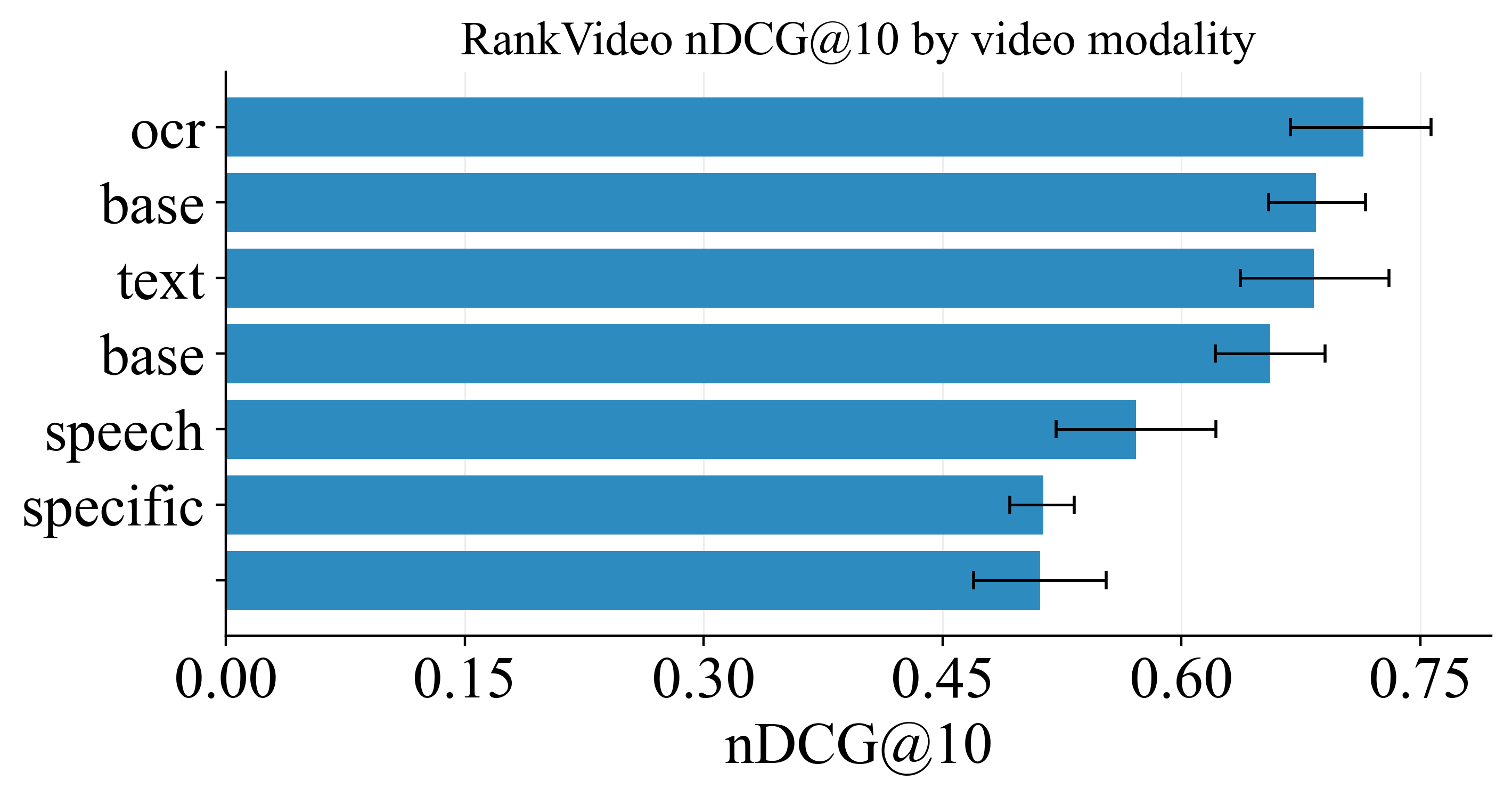}
  \caption{\modelname nDCG@10 by query/video modality. OCR=queries written from text visible in video frames; Text=queries written using only the YouTube description; Base=Wikipedia-title-style queries; Speech=queries written from spoken content; Specific=queries targeting fine-grained aspects of events. Only attributes with $\ge$ 30 test queries are included.}
  \label{fig:rv-ndcg-modality}
\end{figure}

To better understand whether our gains are concentrated in a subset of the evaluation distribution, we break down \textsc{RV} retrieval performance across several metadata slices. Specifically, we report per query nDCG@10 aggregated by \textit{video language} (\autoref{fig:rv-ndcg-lang}), \textit{query event type} (\autoref{fig:rv-ndcg-event}), \textit{video type} (\autoref{fig:rv-ndcg-vtype}), and \textit{video modality} (\autoref{fig:rv-ndcg-modality}).


\section{Disconnect Between Binary Classification and Reranking}
\label{append:disconnect}

\begin{table}[]
    \centering
    \begin{tabular}{c|cccc}
    \toprule
        Method & Accuracy & R@10 & nDCG@10 \\
    \midrule
    \rowcolor{gray!30}
        OE & $-$ & 0.52 & 0.50\\
    \midrule
        Neg-1 & 0.74 & \red{0.48} & 0.52 \\
        Neg-2 & 0.76 & 0.52 & \underline{0.54} \\
        Neg-3 & 0.79 & 0.54 & \underline{0.54} \\
    \bottomrule
    \end{tabular}
    \caption{Evidence of the disconnect between accuracy and reranking performance. Neg: negative result while training \modelname.}
    \label{tab:disconnect}
\end{table}
One failure mode we observed during our training is a disconnect between binary classification\footnote{In this case, yes/no judgments for the relevance of a query-video pair.} and reranking quality. During development, multiple models achieved strong accuracy and recall (\autoref{tab:disconnect}) when evaluated as relevance classifiers. However, these metrics are not indicators of strong second-stage results because they largely reflect performance on easy negatives. 

Classifiers blindly trained to increase accuracy end up being insensitive to the error regime that dominates second-stage retrieval. When reranking, the model is only given the top-k candidates returned by a first-stage retriever, where negatives are hard---semantically or visually plausible near-misses---and therefore disproportionately likely to become high-scoring false positives. While accuracy and reranking performance are slightly positively correlated, we observe that higher accuracy does not always mean strong reranking results. 

To address this, we designed our training supervision to better match the reranking objective by mining negatives from each query's candidate pool and filtering them using teacher confidence. Specifically, we removed likely false negatives by dropping suspected positives. We thresholded suspected positives as non-qrels candidates that the teacher labeled as relevant with high margin. Among the remaining candidates, we retained trusted negatives that the teacher rejected with a large negative margin, and ambiguous negatives that remain difficult but are flagged with lower confidence. We required at least one trusted negative per query. This curation shifts learning away from separating irrelevant videos and toward suppression of retrieval hard negatives while avoiding noisy supervision from unlabeled positives.

\section{Reasoning Examples}
\label{append:examples}

\begin{figure*}[t]
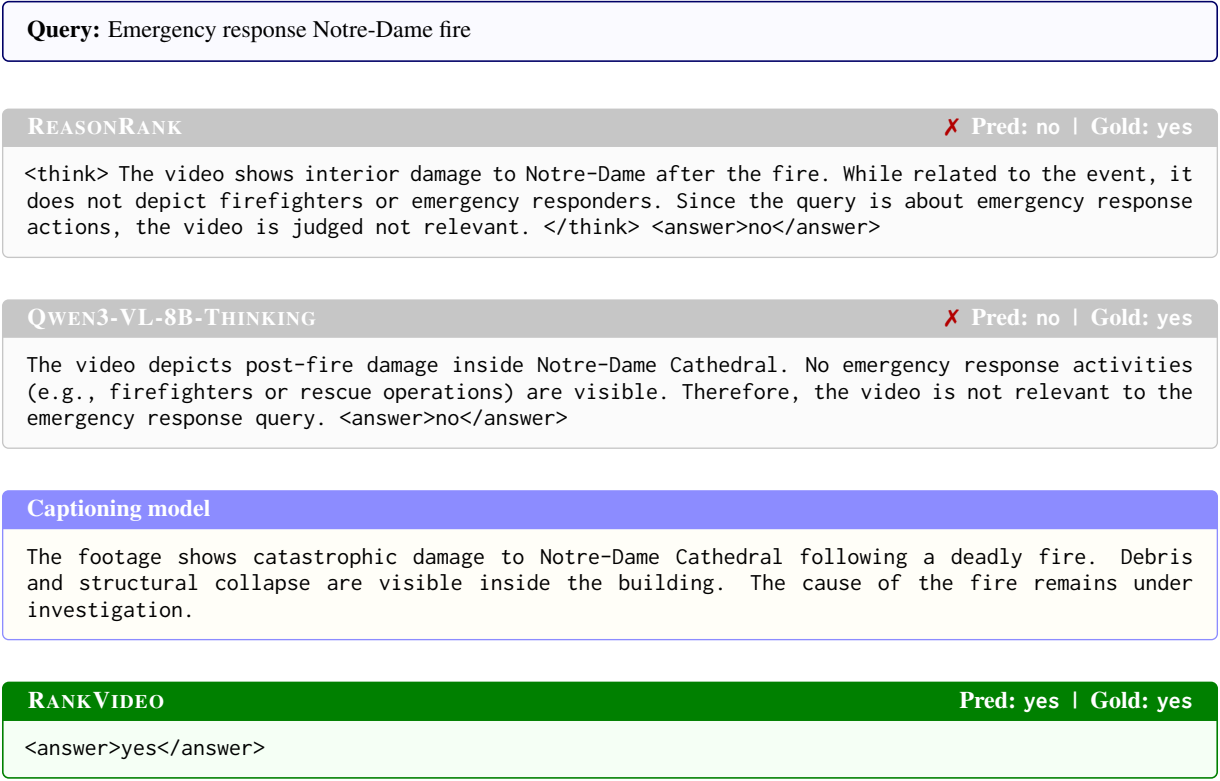

\centering

\setlength{\fboxsep}{6pt}

\begin{tcolorbox}[
  colback=blue!2,
  colframe=blue!40!black,
  boxrule=0.6pt,
  arc=2pt,
  left=6pt,right=6pt,top=4pt,bottom=4pt
]
\small
\textbf{Query:} Emergency response Notre-Dame fire
\end{tcolorbox}

\vspace{4pt}

\begin{tcolorbox}[
  breakable,
  colback=gray!3,
  colframe=gray!45,
  boxrule=0.5pt,
  arc=2pt,
  left=6pt,right=6pt,top=4pt,bottom=4pt,
  title={\small \textsc{ReasonRank} \hfill \xmark\; Pred: \texttt{no} \;|\; Gold: \texttt{yes}},
  fonttitle=\bfseries
]
\footnotesize
\texttt{<think> The video shows interior damage to Notre-Dame after the fire. While related to the event, it does not depict firefighters or emergency responders. Since the query is about emergency response actions, the video is judged not relevant. </think>
<answer>no</answer>}
\end{tcolorbox}

\vspace{2pt}

\begin{tcolorbox}[
  breakable,
  colback=gray!3,
  colframe=gray!45,
  boxrule=0.5pt,
  arc=2pt,
  left=6pt,right=6pt,top=4pt,bottom=4pt,
  title={\small \textsc{Qwen3-VL-8B-Thinking} \hfill \xmark\; Pred: \texttt{no} \;|\; Gold: \texttt{yes}},
  fonttitle=\bfseries
]
\footnotesize
\texttt{The video depicts post-fire damage inside Notre-Dame Cathedral. No emergency response activities (e.g., firefighters or rescue operations) are visible. Therefore, the video is not relevant to the emergency response query.
<answer>no</answer>}
\end{tcolorbox}

\vspace{2pt}

\begin{tcolorbox}[
  breakable,
  colback=yellow!3,
  colframe=blue!45,
  boxrule=0.5pt,
  arc=2pt,
  left=6pt,right=6pt,top=4pt,bottom=4pt,
  title={\small Captioning model},
  fonttitle=\bfseries
]
\footnotesize
\texttt{The footage shows catastrophic damage to Notre-Dame Cathedral following a deadly fire. Debris and structural collapse are visible inside the building. The cause of the fire remains under investigation.}
\end{tcolorbox}

\vspace{2pt}

\begin{tcolorbox}[
  colback=green!4,
  colframe=green!50!black,
  boxrule=0.6pt,
  arc=2pt,
  left=6pt,right=6pt,top=4pt,bottom=4pt,
  title={\small \modelname \hfill \cmark\; Pred: \texttt{yes} \;|\; Gold: \texttt{yes}},
  fonttitle=\bfseries
]
\footnotesize
\texttt{<answer>yes</answer>}
\end{tcolorbox}

\vspace{2pt}

\caption{Reasoning traces for the query \emph{``Emergency response Notre-Dame fire''}. ReasonRank and Qwen3-VL-8B-Thinking models are substantially more verbose then our model. Our final model correctly predicts relevance. Video ID: 45391}
\label{fig:no-reason-trace}
\end{figure*}

In \autoref{fig:no-reason-trace} and \autoref{fig:yes-reason-trace} we provide examples of reasoning traces from the baseline and \modelname rerankers. \autoref{fig:no-reason-trace} shows an example where \modelname did not need to reason to produce a correct answer with a query-video pair, while \autoref{fig:yes-reason-trace} shows an example where \modelname needed to reason. In the example with a reasoning trace, the large difference in reasoning length between \textsc{Qwen3-VL-8B-Thinking} and \modelname is evident.

\section{Use of AI Assistants}
AI assistants were used to improve the fluency of the writing and in some of the code development. 
\clearpage

\begin{figure*}[t]
\centering

\setlength{\fboxsep}{6pt}

\begin{tcolorbox}[
  colback=blue!2,
  colframe=blue!40!black,
  boxrule=0.6pt,
  arc=2pt,
  left=6pt,right=6pt,top=4pt,bottom=4pt
]
\small
\textbf{Query:} Super Bowl 2023 Philadelphia Eagles
\end{tcolorbox}

\vspace{4pt}
\begin{tcolorbox}[
  breakable,
  colback=gray!3,
  colframe=gray!45,
  boxrule=0.5pt,
  arc=2pt,
  left=6pt,right=6pt,top=4pt,bottom=4pt,
  title={\small \textsc{ReasonRank} \hfill \; Pred: \texttt{no} \;|\; Gold: \texttt{no}},
  fonttitle=\bfseries
]
\footnotesize
\texttt{<think> Okay, let me try to figure this out. The query is about the Super Bowl 2023 Philadelphia Eagles. So, the user wants to know if the video is relevant to that. Looking at the video content, it starts with a close-up of Richard Sherman from the 49ers. Then there's info about his salary from 2020-2022. Next, there's game footage of him tackling someone in a yellow uniform (maybe another team?). Then he's celebrating on the field. The video ends with a blue screen and social media stuff. Now, the query mentions Philadelphia Eagles and Super Bowl 2023. The video mostly talks about Richard Sherman's salary and some game play. There's no mention of the Eagles or the 2023 Super Bowl. The focus seems to be on Sherman's career stats and a play. Since the query is specifically about the Eagles in the 2023 Super Bowl, and the video doesn't cover that, I think it's not relevant. Unless maybe the Eagles are involved in the footage, but the caption says Sherman tackles someone in yellow, so probably not Eagles. So the answer should be no. </think> <answer> no </answer>}
\end{tcolorbox}

\vspace{2pt}
\begin{tcolorbox}[
  breakable,
  colback=gray!3,
  colframe=gray!45,
  boxrule=0.5pt,
  arc=2pt,
  left=6pt,right=6pt,top=4pt,bottom=4pt,
  title={\small \textsc{Qwen3-VL-8B-Thinking} \hfill \; Pred: \texttt{no} \;|\; Gold: \texttt{no}},
  fonttitle=\bfseries
]
\footnotesize
\texttt{Got it, let's check the query: Super Bowl 2023 Philadelphia Eagles. Now, look at the video content. The video shows a player in a San Francisco 49ers uniform (jersey number 20, helmets with 49ers logo), social media prompts, salary info for 2020-2022, game action with 49ers, and other teams like Eagles? Wait, let's see. The video has players from the 49ers, and there's a part where the team is in red and gold, which is 49ers. The query is about Super Bowl 2023 Philadelphia Eagles, but the video seems to focus on the 49ers, not the Eagles. Also, Super Bowl 2023 isn't actually a thing yet (2023 would be Super Bowl LVII, but the Eagles' Super Bowl is like 2023 vs the Chiefs maybe? Wait, but the video doesn't mention Philadelphia Eagles or Super Bowl 2023. Let's check all segments. The video starts with a 49ers player, then salary info, then a game action (49ers), then another team (Patriots vs Seahawks?), then a thank you screen. There's no mention of Philadelphia Eagles or Super Bowl 2023. The query is about Eagles in 2023 Super Bowl, but the video is about 49ers, so it's not relevant.
<answer>no</answer>}
\end{tcolorbox}

\vspace{2pt}

\begin{tcolorbox}[
  breakable,
  colback=yellow!3,
  colframe=blue!45,
  boxrule=0.5pt,
  arc=2pt,
  left=6pt,right=6pt,top=4pt,bottom=4pt,
  title={\small Captioning model},
  fonttitle=\bfseries
]
\footnotesize
\texttt{A San Francisco 49ers football player in his white jersey with the number 20 is shown on the field. The video then displays a financial breakdown of his contract for the years 2020, 2021, and 2022, detailing his base salary, guaranteed amounts, bonuses, and total salary-cap hits. Action shots from games are shown, including a player celebrating with arms outstretched. The video concludes with an outro screen thanking viewers and asking them to subscribe.}
\end{tcolorbox}

\vspace{2pt}

\begin{tcolorbox}[
  colback=green!4,
  colframe=green!50!black,
  boxrule=0.6pt,
  arc=2pt,
  left=6pt,right=6pt,top=4pt,bottom=4pt,
  title={\small \modelname \hfill \cmark\; Pred: \texttt{no} \;|\; Gold: \texttt{no}},
  fonttitle=\bfseries
]
\footnotesize
\texttt{The video is not relevant to the query. The query asks about Super Bowl 2023 and the Philadelphia Eagles, but the video content shows a San Francisco 49ers player (number 20) in a Super Bowl LVII jersey, along with salary cap information for the years 2020-2022, and other football-related images. There is no mention or depiction of the Philadelphia Eagles or Super Bowl 2023 specifically.
<answer>no</answer>}
\end{tcolorbox}

\vspace{2pt}

\caption{Reasoning traces for the query \emph{``Super Bowl 2023 Philadelphia Eagles''}. In this instance, all models correctly classify the video as not relevant but we see ReasonRank and Qwen3-VL-8B-Thinking use substantially more tokens then \modelname. Video ID: 45391}
\label{fig:yes-reason-trace}

\end{figure*}

\clearpage
\begin{figure*}[t]
\noindent\fbox{%
    \parbox{.98\textwidth}{%
{\tt
\textbf{[System Prompt]}
You are a helpful assistant specialized in video and text understanding. Given a video, your task is to produce an accurate caption. Respond within <think></think>

\textbf{[User Prompt]}
Caption this video. Respond within <think></think>.
}
}}
\caption{\modelname prompt for stage 1}
\label{prompt:stage1_prompt}
\end{figure*}

\begin{figure*}[t]
\noindent\fbox{%
    \parbox{.98\textwidth}{%
{\tt
\textbf{[System Prompt]}
You are a helpful assistant specialized in video and text understanding. Given a text query and a video, your task is to determine if the video is relevant to the query. Respond with <answer>yes</answer> if the video is relevant, or <answer>no</answer> if it is not.

\textbf{[User Prompt]}
Query: \{query\} Is the video relevant to the query? Respond with <answer>yes</answer> or <answer>no</answer>.
}
}}
\caption{\modelname prompt for stage 2}
\label{prompt:method_prompt}
\end{figure*}

\begin{figure*}[t]
\noindent\fbox{%
    \parbox{.98\textwidth}{%
{\tt
\textbf{[System Prompt]}
You are a helpful assistant specialized in video and text understanding. Given a text query and a video, your task is to determine if the video is relevant to the query. Respond with <answer>yes</answer> if the video is relevant, or <answer>no</answer> if it is not.

\textbf{[User Prompt]}
Query: \{query\} Is the video relevant to the query? Respond with <answer>yes</answer> or <answer>no</answer>.
}
}}
\caption{Prompt for the QwenVL Models}
\label{prompt:qvl}
\end{figure*}

\begin{figure*}[t]
\noindent\fbox{%
    \parbox{.98\textwidth}{%
{\tt
\textbf{[System Prompt]}
You are RankLLM, an intelligent assistant that can rank passages based on their relevance to the query. Given a query and a passage list, you first thinks about the reasoning process in the mind and then provides the answer (i.e., the reranked passage list). The reasoning process and answer are enclosed within <think> </think> and <answer> </answer> tags, respectively, i.e., <think> reasoning process here </think> <answer> answer here </answer>.
}
}}
\caption{Prompt for the \reasonrank}
\label{prompt:reasonrank}
\end{figure*}

\end{document}